\documentclass{emulateapj}

\newcommand{\qa}{SDSS J1030$+$0524 }
\newcommand{\qb}{SDSS J1148$+$5251 }
\newcommand{\lya}{Ly$\alpha$ }
\newcommand{\lyb}{Ly$\beta$ }
\newcommand{\lyg}{Ly$\gamma$ }
\newcommand{\qans}{SDSS J1030$+$0524}
\newcommand{\qbns}{SDSS J1148$+$5251}
\newcommand{\lyans}{Ly$\alpha$}

\newcommand{\lygns}{Ly$\gamma$}

\def\myputfigure#1#2#3#4#5%
{\hskip0.03\textwidth\vskip#5pt
\makebox[0pt]{\hskip#2in
\includegraphics[width=#3\textwidth]{#1}}\vskip#4pt\hfill}


\lefthead{Oh \& Furlanetto}
\righthead{How Universal is the Gunn-Peterson Trough at $\lowercase{z}\sim6$ ?}

\begin{document}

\title{How Universal is the Gunn-Peterson Trough at
$\lowercase{z}\sim6$?: \\ A Closer Look at the Quasar SDSS J1148+5251}

\author{S. Peng Oh\altaffilmark{1} \& Steven R. Furlanetto\altaffilmark{2}}

\altaffiltext{1}{Department of Physics, University of California, Santa Barbara, CA 93106; email: peng@physics.ucsb.edu}
\altaffiltext{2}{Theoretical Astrophysics, Mail Code 130-33, Caltech, Pasadena, CA 91125; email: sfurlane@tapir.caltech.edu}
\vspace{\baselineskip}
\begin{abstract}
Detectable flux is visible in the \lya and \lyb troughs of the highest redshift ($z=6.42$) quasar found to date, \qbns.  This has previously been interpreted as continuum contamination from an interloper galaxy at $z=4.94$. We examine the \lyg trough of \qb and show that this interpretation is untenable: the spectrum does not show the continuum break in a $z=4.94$ galaxy expected from absorption by the
intervening Ly$\alpha$ forest. Therefore, flux must be leaking through at least one of the troughs from the quasar itself. Contrary to previous claims, the flux ratios in the Ly$\alpha$ and \lyb troughs are consistent with pure transmission. From the \lyg trough, we place an {\it upper} bound on the effective Ly$\alpha$ optical depth at $z\sim 6.2$ of $\tau_{\rm eff} < 14.3 \, (2\sigma)$. This implies a highly ionized IGM along this line of sight and significant cosmic variance in the transition toward complete Gunn-Peterson absorption. Detailed study of the observed transmission features will shed light on this era.  
\end{abstract}

\section{Introduction}
\label{sec:introduction}

Despite spectacular progress in recent years, the reionization history of the universe remains shrouded in mystery. The high electron scattering optical depth observed by WMAP indicates that reionization
may have begun as early as $z\sim 20$ (Kogut et al 2003; Spergel et al 2003). On the other hand, spectra of high redshift quasars indicate a sharp change in the neutral hydrogen fraction and ionizing
background at $z\sim6$, suggesting this era marks the end of the reionization epoch (Becker et al 2001, Fan et al 2002). Taken in tandem, these observations suggest a highly complex reionization history.

A crucial missing piece of the puzzle is the precise ionization state of the IGM at $z=6$. The large optical depth of the IGM to hydrogen Lyman transitions implies that Gunn \& Peterson (1965) absorption can (at best) constrain the volume and mass-weighted neutral fractions to be $x_{\rm HI,V} \ge 10^{-3},\, x_{\rm HI,M}\ge 10^{-2}$ respectively (Fan et al 2002). It is plausible that the IGM is still highly ionized at $z=6$,
and with full reionization occurring much earlier.  On the other hand, modeling of the spectral regions around two quasars at $z \ga 6.2$ may indicate larger neutral fractions ($x_{\rm HI} \ga 0.2$; Wyithe \& Loeb 2004a; Mesinger \& Haiman 2004). At stake is the very nature of reionization: if these claims are correct, then reionization must be an extremely rapid process, since the IGM is known to be highly ionized ($x_{\rm HI} \le 10^{-5}$) by $z=5.9$ along all observed lines of sight (Fan et al 2003).

It is thus worth considering the IGM absorption in more detail.  In particular, a completely dark Gunn-Peterson trough at $z\sim6$ is not necessarily universal.   Both the \lya and \lyb troughs of the highest redshift quasar found to date, \qbns, contain detectable flux (White et al 2003). In this {\it Letter}, we show that its previous interpretation as continuum contamination from an interloper galaxy at $z=4.94$ is inconsistent with the observed flux ratios blueward and redward of the expected continuum break due to \lya forest absorption. Therefore, some of the flux is true transmission due to holes in the high redshift \lya forest. We derive our strongest contraints from the \lyg trough, which was ignored in previous analyses and should be relatively uncontaminated by flux from an interloper. The presence of flux transmission implies that either the IGM is still highly ionized at $z=6$ or that there is significant cosmic variance in the reionization epoch along different lines of sight.

\section{The Lyman Gamma Trough in SDSS J1148+5251}
\label{section:lyg_trough}

Detectable flux and a network of transmission features is seen in both the \lya and \lyb troughs of \qbns.  White et al (2003) interpreted this to be continuum contamination from interloper galaxies at $z=4.94$ for two reasons: (i) strong \ion{C}{4} absorption features appear at $z=4.9$, so the apparent strong \lyb spikes at $z=6.03, 6.06$ could just be \lya emission from $z \approx 4.94$. (ii) The flux ratios of the troughs appear inconsistent: there is too little light in the \lyb trough given the \lya transmission. Although the {\it a priori} probability of such an interloper is small, it increases if the interloper also lenses \qbns, boosting its probability of detection.

By examining the \lyg trough of \qbns, we argue that such an interpretation is untenable. We first note that if residual flux in {\it any} of the Ly$\alpha,\beta,\gamma$ troughs can be attributed to the quasar rather than an interloper galaxy, the IGM must still be highly ionized along the line of sight.  Consider the boundary of the quasar proximity zone at $z_{\rm HII}\approx 6.33$, where the \lyans, \lyb absorption increase rapidly. {\it There is also a sharp jump in absorption at the boundary of the \lyg trough, strongly suggesting that the transmission features redward of the \lyg trough are genuine}. Even if we exclude the sharp transmission spike at $\lambda=7205$ \AA \ (putatively a \lya emission line from a galaxy at $z=4.94$), the residual flux changes from $F_{-20}(z_{\gamma}=6.33-6.40)=29.0\pm1.5$
(within the quasar proximity zone) to $F_{\-20}(z_{\gamma}=6.25-6.32)=4.3\pm1.4$ (outside the quasar proximity zone), where $F_{-20}$ is the flux $F$ in units of $10^{-20} {\rm erg \, s^{-1} \, cm^{-2}}$ \AA$^{-1}$.  It would be remarkable if the protocluster at $z\approx5$ lined up so as to exactly coincide with the quasar proximity zone in \lygns: it is much more likely that the sudden change is due to genuine \lyg absorption.  This suggests the flux seen in the \lyb forest at $z>5.95$, before the onset of \lyg absorption, represents true transmission. 

It is also crucial that there is detectable flux in the \lyg trough. Consider the wavelength stretch corresponding to $z=6.17-6.32$ in the various absorption troughs. This lies outside the quasar's apparent region of influence and ends where the \lyg trough becomes contaminated by Ly$\delta$ and higher order absorption ($\lambda=6962$ \AA, corresponding to $z_{\gamma}=6.16$).   White et al (2003) suggest that the \lyans, \lyb troughs are contaminated by continuum emission from the interloper galaxy so that detected flux is not significant. To be conservative, let us ignore the transmitted flux in the \lyb trough, which is potentially also contaminated by emission line features from the $z\approx5$ proto-cluster. We find that $F_{-20}$(\lyans)$=3.0 \pm0.9$ while $F_{-20}$(\lygns)$=4.9 \pm0.9$; these regions correspond to $\lambda=1467-1498$ \AA \ and $\lambda=1174-1199$ \AA \ in the rest frame of an interloper galaxy at $z=4.94$. {\it These flux ratios are strongly inconsistent with the expected continuum of a $z\approx5$ galaxy, which should have a strong break due to the intervening \lya forest}. Songaila \& Cowie (2002) obtain a flux suppression factor $T_{\alpha}=0.14 \pm 0.03$ for the \lya forest in this redshift and wavelength interval (Becker et al 2003 find $T_{\alpha}=0.11$).  Even if the flux in the \lya trough is entirely due to continuum contamination from the interloper, the implied continuum contribution to the \lyg trough would be $F_{-20}=0.4 (T_{\alpha}/0.14) \ll F_{-20}$(\lygns), which lies well within the noise. Thus, almost all the observed flux in the \lyg trough must be genuine transmitted flux from the quasar. Because it is almost completely free from continuum contamination, the \lyg trough places the strongest constraint on the IGM ionization state along this line of sight.

\begin{deluxetable*}{ccrrrrrlll}
\tabletypesize{\scriptsize}
\tablecolumns{7}
\tablewidth{0pc}
\tablecaption{Absorption Optical Depths from $z=6.17-6.32$ in \qb}
\tablehead{
\colhead{Trough} & \colhead{Flux\tablenotemark{a} }&
\colhead{Continuum\tablenotemark{a,b}} & \colhead{Total Optical Depth} &
\colhead{Line Optical Depth\tablenotemark{c}} & \colhead{$\tau_{\alpha}(\beta=2)$\tablenotemark{d}} 
& \colhead{$\tau_{\alpha}(\beta=3)$\tablenotemark{d}} 
}
\startdata
\lya & $3.0 \pm 0.9$ & (2560,1730,1690) & $(6.8,6.4,6.4)\pm 0.3$ & $(6.8,6.4,6.4)\pm 0.3$ & $(6.8,6.4,6.4)\pm 0.3$ & $(6.8,6.4,6.4)\pm 0.3$\\
\lyb & $9.0 \pm 1.1$ & (2140,1870,1710) & $(5.5,5.3,5.2)\pm 0.1$ & $(3.1,3.0,2.9)\pm 0.3$ & $(8.2,7.8,7.6)\pm 0.9$ & $(6.7,6.4,6.2)\pm 0.7$\\
\lyg & $4.9 \pm 0.9$ & (2070,1910,1710) & $(6.1,6.0,5.9)\pm 0.3$ & $(2.5,2.4,2.3)\pm 0.3$ & $(11.5,11.2,10.7)\pm 1.4$ & $(8.4,8.2,7.8)\pm 1.0$
\tablenotetext{a}{ In units of $10^{-20} {\rm erg \, s^{-1} \, cm^{-2} \AA^{-1}}$.}
\tablenotetext{b}{Derived from the LBQS composite spectrum, and the Telfer et al 2002 spectral indices for radio-quiet ($\alpha_{\rm EUV}=-1.57$) and radio-loud ($\alpha_{\rm EUV}=-1.96$) quasars respectively. Subsequent triplets for the calculated optical depths reflect this ordering.}
\tablenotetext{c}{After subtracting off the contribution from foreground absorption. For \lyb trough: $\tau_{\rm fg}=\tau_{\alpha}(z=5.1)=2.38\pm0.32$; for \lyg trough: $\tau_{\rm fg}=\tau_{\alpha}(z=4.8)+\tau_{\beta}(z=5.9)=(1.95\pm0.25)+(1.59 \pm 0.02)=3.54\pm0.25$.}
\tablenotetext{d}{The effective equivalent \lya optical depth inferred from the nonlinear conversion described in the text, assuming $\tau(\Delta) \propto \Delta^{\beta}$.}
\enddata
\end{deluxetable*}

Because there are indisputably \ion{C}{4} absorbers at $z=4.9$, the putative protocluster may conceivably have another effect: it could ionize the IGM at $z=5$, creating a transparent Ly$\alpha$ window there. 
However, the proximity zone of the interloper galaxy itself is too small. We need to highly ionize the IGM on comoving lengthscales $L \approx 55 (\Delta z/0.1)$ Mpc at $z=5$. If the flux in the Ly$\alpha$ trough is interpreted as the continuum of the interloper, then
$F_{-20}(\lambda=1450 \AA)=3.9$. If we take $F(\lambda=1450 \AA)/F(\lambda=900 \AA)=4.6$ as is observed in $z=3.4$ LBGs (Steidel et al 2001) for the strongest ionizing continuum possible (corresponding to a $\sim 100\%$ escape fraction for ionizing photons), then we infer $L_{\nu}(1 {\rm Ryd})=1.4 \times 10^{28} {\rm erg \, s^{-1} \, cm^{-2} \, Hz^{-1}}$, roughly implying a star formation rate ${\rm SFR} \sim 14 {\rm M_{\odot} yr^{-1}} $ for a Salpeter IMF. Assuming the mean background ionizing rate in units of $10^{-12} {\rm s}^{-1}$ to be $\Gamma_{-12} (z=5)\approx 0.15$ (e.g., Fig 2 of Fan et al 2002), the local radiation field becomes comparable to the metagalactic ionizing at $r\approx 1.3$ Mpc comoving, far too small to be of interest.  Lyman-break galaxies (LBGs) at $z\approx 3-4$ also show excess Ly$\alpha$ transmission on lengthscales $L\approx 0.5$ Mpc comoving (Adelberger et al 2003), roughly corresponding to the distance a $600 \, {\rm km \, s^{-1}}$ wind would travel on the $\sim 300$ Myr star formation timescale of LBGs. It is unlikely a $z=5$ galaxy would affect a significantly larger volume. 
Galaxies near (but not along) the line of sight could contribute additional ionizing photons. Note, however, that we require a factor $\sim 5$ increase in $T_{\alpha}$ over a lengthscale $\sim 10$ times larger than the comoving correlation length for highly biased galaxies. There is no evidence for such strong fluctuations in transmission at this redshift: typically $\sigma_{\rm T\alpha}/T_{\alpha}\sim 0.3$ (Songaila \& Cowie 2002).  In any case, an anomalously large protocluster should be easily visible in narrowband optical searches.

\section{Implications for the IGM at $z=6$}
\label{section:tau}

Having established the reality of residual flux from \qbns, we will now consider its implications for reionization at $z \sim 6$.  We must first estimate the effective optical depths $\tau_{\rm eff}$ in each transition.  We begin by summarizing our methodology and the error budget (which is frequently underestimated).

The errors for Ly$\gamma$ transmission $T_{\gamma}=T_{\rm tot}/(T_{\alpha}T_{\beta})$ must include not only pixel noise but also the cosmic variance in foreground transmission in Ly$\alpha$ and
Ly$\beta$. Previous analyses have often neglected this additional scatter.  It has been measured in quasar samples; we use the values in Table 2 of Songaila \& Cowie (2002), who tabulate the measured
$\langle T_\alpha \rangle, \sigma_{\rm T\alpha}$ for 6 redshift bins between $z=4.1-5.5$. However, these use fixed wavelength intervals, and $\sigma_{\rm T\alpha}$ obviously depends on the corresponding comoving length $L$. We cannot simply assume Poisson statistics, because long wavelength modes could dominate the variance. To estimate $\sigma_{\rm T\alpha}$ we follow the {\it ansatz} of Lidz et al (2002), who argue that the transmission power spectrum takes the shape (though not the normalization) of the linear mass power spectrum on large scales, yielding:
\begin{equation}
\sigma_{\rm T\alpha}^{2}=2 \int_{0}^{\infty} \frac{dk}{2 \pi} \left[ \frac{{\rm sin}(kL/2)}{kL/2}\right]^{2} P_{f}(k) + \frac{\sigma_{n}^{2}}{N},
\label{eqn:sigmaF}
\end{equation} 
where $\sigma_{n}$ is the noise per pixel, $N$ is the number of pixels, and $P_{f}(k)=B{\rm exp}(-ak^{2})\int_{k}^{\infty}(dk/2\pi)k P_{\rm mass}(k)$, where $a\approx k_{\rm J}^{-1/2}$, $k_{J}$ is the Jeans wavenumber, and $B$ is normalized to the observed $\sigma_{\rm T\alpha}$ for some observed
stretch $L_{\rm obs}$. The first term in equation (\ref{eqn:sigmaF}) typically dominates by an order of magnitude. For the large lengthscales of interest, $P_{f}(k)$ is fairly flat, and we have approximately $\sigma_{\rm T\alpha}^{2} \propto L^{-1}$, as expected for a white noise power spectrum. Note that this {\it ansatz} for the transmission power spectrum $P_{f}(k)$ assumes a homogeneous ionizing background $\Gamma$; if $\Gamma$ exhibits small-scale fluctuations, $\sigma_{\rm T\alpha}$ could be larger.  However, there is {\it no} cosmic variance in $T_\beta$, because we have observed the 
corresponding \lya transmission (at $z_\alpha=5.9$) along the same line of sight. The main uncertainty comes from the optical depth conversion between different lines (see below). 

Another crucial item is uncertainty in the quasar continuum. This is particularly important in comparing relative absorption between the different troughs. The Ly$\alpha$ line is much broader than the Ly$\beta$ and Ly$\gamma$ lines and generally spills over into the observed trough. Uncertainty in the strength of the line creates significant uncertainty in the inferred $\tau_{\alpha}$. A simple way to quantify this is to compare the optical depths inferred assuming the composite spectrum from the Large Bright Quasar Survey (LBQS; Brotherton et al 2001) and the far-UV quasar power law spectrum of Telfer et al (2002). The former includes all the emission line structure, while the latter assumes a pure power law; the difference between the two therefore captures the uncertainty in emission line contribution.  We normalize the continuum to the observed flux at $\sim 1290 {\rm \AA}$. 

Table 1 lists the measured effective line optical depths $\tau_i \equiv \tau_{\rm eff}$(\lyans) and their associated $1\sigma$ errors.  The results for the \lya and \lyb line optical depths are comparable to those of White et al. (2003).  They argued that the pair are incompatible with each other, because $\tau_{i} \propto \lambda_{i} f_{i}$ (where $f_{i}$ is the oscillator strength of line $i$), implying $\tau_{\alpha}/\tau_{\beta}=6.24$ and $\tau_{\alpha}/\tau_{\gamma}=17.93$. However, these relations are only true at fixed density, and only apply if the IGM is homogeneous. For the large comoving lengths which we are considering, the flux transmission comes from a variety of densities:
\begin{equation}
\langle T_{i} \rangle = \langle {\rm exp}\left( -\tau_{{\rm eff},i} \right) \rangle = \int {\rm exp}\left[ -\tau_{i}(\Delta) \right] P(\Delta) d\Delta,
\label{eqn:flux_transmission}
\end{equation}
where $P(\Delta)$ is the probability distribution of overdensities $\Delta \equiv \rho/\bar{\rho}$, $\tau_{i} \propto \lambda_{i} f_{i} (1+z)^{4.5} \Delta^{2} \alpha(T)/\Gamma$ (e.g., Hui \& Gnedin 1997) in the optically thin limit, and $\alpha(T)$ is the recombination coefficient.  We use the form of $P(\Delta)$ given by Miralda-Escud\'e, Haehnelt \& Rees (2000), which is a good fit to numerical simulations. If we assume an equation of state $T\propto \Delta^{\gamma}$ (where $\gamma \sim 0$--$1$) and a fluctuating radiation field whose amplitude may be density dependent, $\Gamma = \Gamma_{o} \Delta^{\xi}$, then $\tau_{i} =A(z) (1+z)^{4.5} \Delta^{\beta}$, where $\beta={2-0.7\gamma-\xi}$. We can solve for the normalization constant $A(z)$ by demanding $\langle T_{i}(A) \rangle=
T_{\rm obs}(z)$. Note that $A(z) \propto f_{i} \lambda_{i}/\Gamma_{o}$. 

The integral in equation (\ref{eqn:flux_transmission}) can be evaluated by the method of steepest descents (Songaila \& Cowie 2002, Songaila 2004): 
\begin{equation}
\tau_{\rm eff}= -A^{1/(1+3\beta/4)}+\frac{0.83}{(\beta+4/3)}{\rm ln}(A) + {\rm const}.
\label{eqn:tau_eff}
\end{equation}
The leading term implies that $\tau_{\rm eff} \propto A^{0.4}$ for $\beta=2$ (corresponding to a uniform radiation field and an isothermal equation of state). Thus, in an inhomogeneous universe the optical depth increases more slowly than linearly with the oscillator strength; conversely, it increases more 
slowly with a weaker radiation field.  Evaluating equation (\ref{eqn:flux_transmission}) numerically, we find that $\tau_{\alpha}/\tau_{\beta} \approx 3$, and $\tau_{\alpha}/\tau_{\gamma} \approx 5-6$, with weak dependence on redshift, the equation of state, and $\Gamma$.  This is because transmission is dominated by rare voids, and the primary effect of decreasing $f_i$ is to increase the range of densities sampled by the line.

This behaviour becomes even more marked if $\Gamma$ is not uniform, which is certainly the case before reionization is complete. As a fiducial case, consider a uniform radiation field $\Gamma=0.05$ at $z=6.15$, which produces a Ly$\alpha$ effective optical depth $\tau_{\alpha}=7$. In this case, $(\tau_{\alpha}/\tau_{\beta},\tau_{\alpha}/\tau_{\gamma})=(2.7,4.9)$. Now let us consider various scenarios that would produce the same $\tau_{\alpha}$ but different $(\tau_{\beta},\tau_{\gamma})$. 
Suppose the optical depth $\tau \propto \Delta^{\beta}$ has $\beta>2$, which could, for example, mimic self-shielding in overdense regions.  In this case, from equation (\ref{eqn:tau_eff}), the optical depth increases even more slowly with oscillator strength: for $\beta=3$, ($\tau_{\alpha}/\tau_{\beta}, \tau_{\alpha}/\tau_{\gamma})=(2.1,3.4)$. Only if the optical depth is independent of overdensity ($\beta=0$) will we recover the linear scaling $\tau_{\rm eff} \propto A$. Fluctuations in the radiation field that are uncorrelated with density fluctuations (so that $\tau \propto \Delta^{2}$) will produce similar effects. For example, a radiation field with a lognormal probability distribution $(\bar{\Gamma},\sigma_{\rm ln \Gamma})=(0.02,1)$ yields  $(\tau_{\alpha}/\tau_{\beta},\tau_{\alpha}/\tau_{\gamma})=(1.9,2.9)$.

Thus, the fluctuating density and radiation fields introduce considerable uncertainty in the relations between $\tau_{\alpha},\tau_{\beta},$ and $\tau_{\gamma}$. The effective optical depth is often used to infer $x_{\rm HI,V}$ and $x_{\rm HI,M}$ in the IGM, a procedure that we caution is fraught with uncertainty. In the case of $x_{\rm HI, M}$, it is almost meaningless. It is easy to see why: we can infer $\langle x_{\rm HI} \rangle \propto \langle \tau \rangle$ from $\langle e^{-\tau} \rangle$ only in the limit where $\tau \ll 1$ and $\langle e^{-\tau} \rangle \approx 1 -\langle \tau \rangle$; otherwise, our ignorance of the full probability distribution $P(\tau)$ means that a wide variety of $\langle x_{\rm HI} \rangle$ would be consistent with a given observed $\langle e^{-\tau})$. As an illustration, we show in Figure~1 the logarithmic integrand $y \Delta P(\Delta) \propto \langle y \rangle$, for various quantities $y$; all the curves have been normalized to have unit area. Since $\langle e^{-\tau} \rangle$ is heavily weighted toward voids and $\langle x_{\rm HI} \rangle$ is weighted toward overdense regions, estimating one from the other is extremely model-dependent.  In particular, $x_{\rm HI,M}$ is heavily weighted toward large overdensities, and the quasar spectra essentially leave it unconstrained. For instance, the neutral fraction could be considerably lower in overdense regions without affecting the transmitted flux. 

\myputfigure{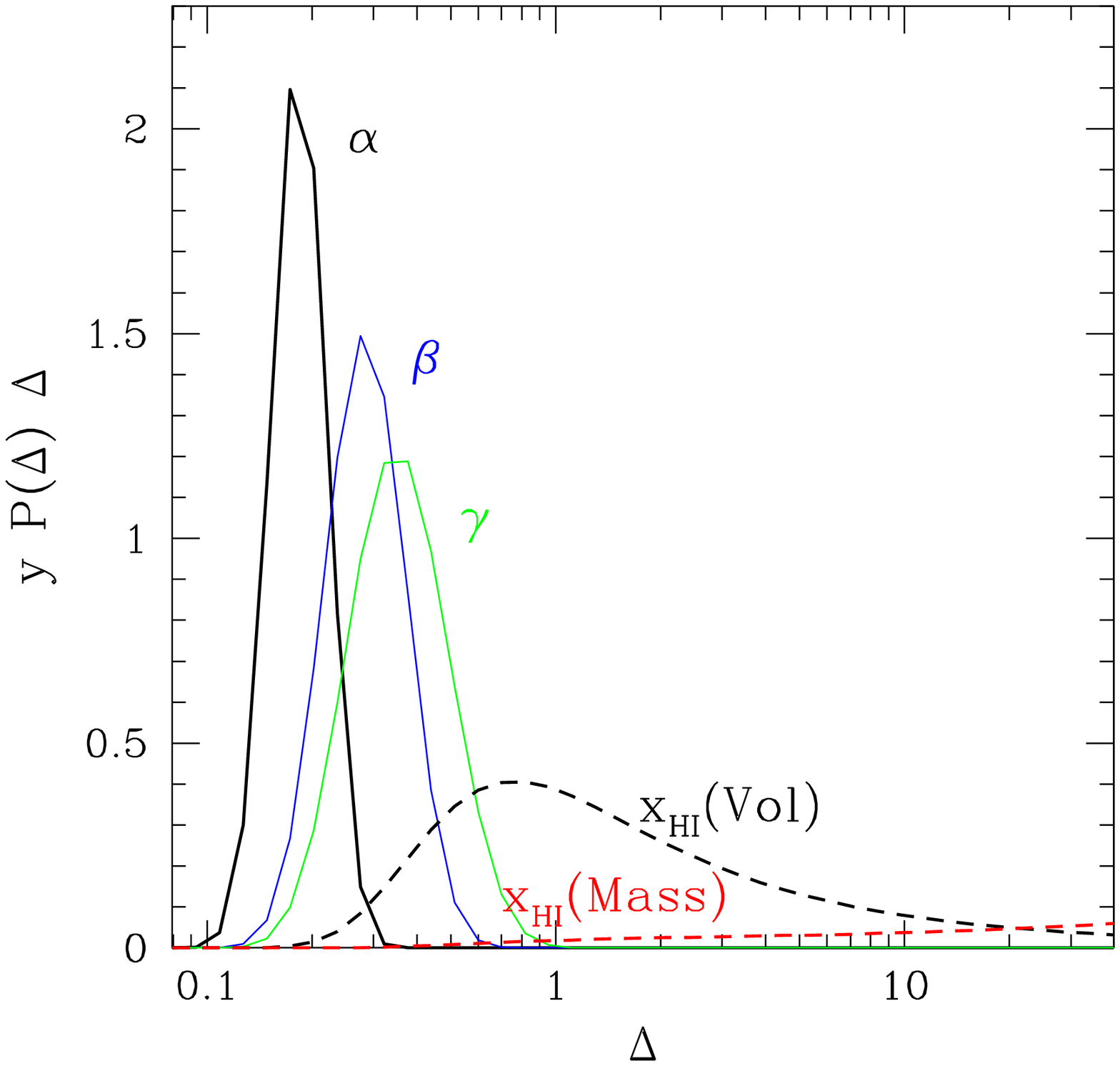}{3.3}{0.5}{-25}{-10}
\figcaption{\label{fig:flux_integrand}
The logarithmic integrand $y \Delta P(\Delta) \propto \langle y \rangle$ as a function of $\Delta$ for $\langle T_{\alpha} \rangle, \langle T_{\beta} \rangle, \langle T_{\gamma} \rangle, \langle x_{\rm HI,V} \rangle, \langle x_{\rm HI,M} \rangle$. We assume a uniform ionizing background $\Gamma= 0.04$ at $z=6.15$ and an isothermal equation of state. The integrand for $\langle x_{\rm HI,M} \rangle$ peaks at high overdensities to the right of the plot. As the overlap integral between the different Lyman series lines drops, the uncertainty in inferring one from the other increases. More importantly, there is considerable uncertainty in inferring $\langle x_{\rm HI} \rangle$ from quasar spectra; in particular, estimates for $\langle x_{\rm HI,M} \rangle$ are almost meaningless.}
\vspace{\baselineskip}

With these caveats, we list $\tau_\alpha$ derived from each of the three transitions in the rightmost columns of Table~1 for $\beta=2,\,3$.  The flux ratios in all 3 troughs are entirely consistent with flux transmission from the quasar without any contamination from an interloper:  the equivalent Ly$\alpha$ optical depths all lie within $1-2\sigma$ of one another. The apparent inconsistency in White et al (2003) appears simply because they did not integrate over the IGM density distribution. Again, we caution that the optical depth conversion between different transitions has large uncertainties (not reflected in the error bars) due to uncertainty in the probability distribution of optical depths $P(\tau)$. The true effective optical depth inferred from the \lya and \lyb troughs could also be somewhat higher if there is some continuum contribution from an interloper, which we cannot completely rule out.  The most stringent optical depth constraint comes from the \lyg trough, which implies $\tau_{\alpha} < 14.3 \ (2 \sigma)$, with a most likely value $\tau_{\alpha} \approx 6$--$10$. By contrast, in \qans, the $1\sigma$ ($2\sigma$) lower limit to the optical depth in the Ly$\beta$ trough is $\tau_{\rm eff} >11.1$ (9.9); the \lyg trough yields similar constraints.

\section{Discussion}

In this {\it Letter}, we have argued that the residual flux in the Ly$\alpha,\beta$, and $\gamma$ troughs of SDSS J1148+5251 is not due to an interloper galaxy but represents true transmission in the $z \ga 6$ IGM. This places an upper bound on the effective \lya optical depth of $\tau_{\rm eff} \le 14.3 \ (2\sigma)$, implying that the IGM is still highly ionized at $z\sim 6.3$. It has been argued that the size of the \ion{H}{2} regions of the two highest redshift quasars $R_{\rm HII} \approx 4.5$ Mpc imply $x_{\rm HI} > 0.1$ in this range (Wyithe \& Loeb 2004a): $R_{\rm HII} \approx 7 x_{\rm HI}^{-1/3} (t_{\rm age}/10^{7} \ {\rm yr})^{1/3}$ Mpc, where $t_{\rm age}$ is the lifetime of the quasar. This is of course strongly dependent on the assumed template spectrum; if the escape fraction of ionizing photons in high-redshift quasars is somehow smaller, the constraint weakens. The quasars could also be lensed and hence intrinsically fainter (Haiman \& Cen 2002), though follow-up HST observations have failed to detect multiple images in either quasar (White et al 2004 in preparation). For our purpose, we note that \ion{H}{2} region radius is determined by the mass-weighted neutral fraction $x_{\rm HI,M}$.  As we showed in \S \ref{section:tau}, this is very poorly constrained by $\tau_{\rm eff}$, because voids containing only a small fraction of the mass dominate the transmission. For instance, if $x_{\rm HI, M} \sim 0.1$ of the baryons are in neutral self-shielded halos, we could still have $x_{\rm HI, V} \sim x_{\rm HI, M}/\delta \sim 0.1/200 \sim 5 \times 10^{-4}$, which could easily be accommodated by our results.

A second argument in favor of $x_{\rm HI} \ga 0.1$ is an indirect detection of the Gunn-Peterson damping wing (Mesinger \& Haiman 2004). There is a stretch at the boundary of the \ion{H}{2} region of SDSS J1030+0524 with Ly$\beta$ transmission but no Ly$\alpha$ transmission, implying $6 <\tau_{\rm eff} < 11$.  The absence of any \lya transmission from low-density voids in this segment implies a source of smooth opacity, attributed to a Ly$\alpha$ damping wing that requires a large optical depth $\tau > 10^{3}$.  However, \qb contains no such transition region. 
We also caution that the statistical significance of such transition regions is still unclear. For instance, in \qb there is a stretch from $z=5.95-6.0$ with \lyb transmission ($F_{-20}=37.9 \pm 1.7$) but no significant \lya flux ($F_{-20}=2.5\pm 1.7$). Again, the optical depth ratios are consistent with pure flux transmission ($\tau_{\alpha}=6.6\pm0.7$, $\tau_{\beta}=2.0\pm 0.4 \Rightarrow \tau_{\rm eff,\alpha}\approx 5.5 \pm 1.2$). This stretch is $\sim 2.5$ times longer than the $\Delta z=0.02$ zone seen in \qans,
over which even damping wing absorption would change substantially; 
in any case, we have argued that the IGM is highly ionized along this stretch. Such regions may therefore be fairly generic and not indicative of a damping wing. More detailed analysis, incorporating a pixel-by-pixel analysis, variance in the foreground \lya forest, and better theoretical modeling of the fluctuating radiation and density fields in realistic models of reionization, would help shed light on this issue.  

Nonetheless, if the region around \qa is significantly neutral, this may be a hint of large cosmic variance in the epoch of reionization.  
Indeed, a strongly fluctuating $\tau_{\rm eff}$ is itself a signature of the pre-overlap era. In the extreme interpretation that the \lya and \lyb troughs of \qa indicate $x_{\rm HI}\sim 0.2$ down to $z=5.95$, this implies large modulation in the ionization fraction and typical bubble sizes of order $\Delta z=0.38$, or $L\sim 150$ Mpc comoving.  This is substantially larger than the \ion{H}{2} regions of these extremely bright and rare quasars, $R_{\rm HII} \sim 30$ Mpc comoving.  It is also larger than theoretical expectations for the scale of typical \ion{H}{2} regions at the tail end of reionization (Furlanetto et al. 2004; Wyithe \& Loeb 2004b), so we consider such a scenario to be extremely unlikely. At the other extreme, the measured $\tau_{\rm eff}$ are compatible with a slightly faster increase in $\tau_{\rm eff}$ along the sightline to \qa ($\tau_{\rm eff} >9.9$, \ 2$\sigma$), than to \qb, ($\tau_{\rm eff}\approx 6$--$10$).   Whether a highly-ionized universe can tolerate such scatter is unclear.  We have shown that a simple $\tau_{\rm eff}$ analysis is a blunt instrument, and more sophisticated interpretation is required to put strong constraints on reionization.

Note that surveys of \lya emitters have found no evolution of the luminosity function of \lya emitters at $z=5.7$ and $z=6.5$ (Malhotra \& Rhoads 2004, Stern et al 2004), which has been interpreted as evidence against percolation taking place at $z\sim 6$, in agreement with our conclusions.

What other signatures of large Ly$\alpha$ optical depth could emerge from the spectra? One possibility is the \ion{O}{1} absorption forest (Oh 2002).  \ion{O}{1} is an excellent tracer of neutral hydrogen: it has a very similar ionization potential $E=13.62$eV, so it lies in tight charge exchange equilibrium. Furthermore, its $1302$ \AA \ absorption line lies redward of the Ly$\alpha$ forest. Although the \ion{O}{1} forest coincides with a noisy portion of the night sky, the sightline to \qa (which may contain substantially neutral regions) has no lines with $W_{\rm obs} > 0.5$ \AA \ and perhaps $\le 3$ lines with $W_{\rm obs} > 0.3$ \AA \ (X. Fan, private communication). Significantly more lines are expected if the universe is substantially neutral (Oh 2002), although this is model-dependent. The \ion{O}{1} forest clearly merits further investigation, though a good standard star calibration is essential to removing strong telluric features from the atmosphere. 

\vspace{.1in}
We thank R. White, X. Fan for providing the spectra of SDSS J1148+5251, and X. Fan for stimulating conversations. We also thank Z. Haiman, A. Lidz, P. Madau, A. Meisinger for helpful discussions/correspondence, and Colleen Schwartz for technical assistance. SPO gratefully acknowledges support by NSF grant AST-0407084, PHY99-07949 and the hospitality of KITP.


\begin{thebibliography}{}
\bibitem[Barkana \& Loeb(2004)]{bl04} Barkana, R., \& Loeb, A. 2004, \apj, 601, 64
\bibitem[Becker et al.(2001)]{becker01} Becker, R. H., et al. 2001, \aj, 122, 2850
\bibitem[Brotherton et al.(2001)]{brotherton01} Brotherton, M.~S., 
Tran, H.~D., Becker, R.~H., Gregg, M.~D., Laurent-Muehleisen, S.~A., \& 
White, R.~L.\ 2001, ApJ, 546, 775
\bibitem[Fan et al.(2002)]{fan02} Fan, X., et al. 2002, \aj, 123, 1247
\bibitem[Fan et al.(2003)]{fan03} Fan, X., et al. 2003, \aj, 125, 1649
\bibitem[Furlanetto, Zaldarriaga, \& Hernquist(2004)]{furl04} 
Furlanetto, S.~R., Zaldarriaga, M., \& Hernquist, L.\ 2004, \apj, 613, 1 
\bibitem[Gunn \& Peterson(1965)]{gunn65} Gunn, J.~E.~\& Peterson, B.~A.\ 1965, \apj, 142, 1633 
\bibitem[Haiman \& Cen(2002)]{zoltan2002} Haiman, Z.~\& Cen, R.\ 
2002, \apj, 578, 702 
\bibitem[Hui \& Gnedin(1997)]{hui97} Hui, L.~\& Gnedin, 
N.~Y.\ 1997, \mnras, 292, 27 
\bibitem[Kogut et al (2003)]{kogut} Kogut A. et al, 2003, ApJS, 148, 161 
\bibitem[Lidz et al.(2002)]{letal02} Lidz, A., Hui, L., Zaldarriaga, M., \& Scoccimarro, R. 2002, \apj, 579, 491
\bibitem[Malhotra \& Rhoads (2004)]{mal_rhoads} Malhotra \& Rhoads, ApJL, in press, astro-ph/0407408
\bibitem[Mesinger \& Haiman(2004)]{meisinger_haiman} Mesinger, A.~\& 
Haiman, Z.\ 2004, ApJ, 611, L69 
\bibitem[Miralda-Escud{\' e}, Haehnelt, \& Rees(2000)]{jordi2000} 
Miralda-Escud{\' e}, J., Haehnelt, M., \& Rees, M.~J.\ 2000, \apj, 530, 1
\bibitem[Oh(2002)]{oh02} Oh, S.~P.\ 2002, \mnras, 336, 1021 
\bibitem[Songaila \& Cowie(2002)]{sc02} Songaila, A., \& Cowie, L. L. 2002, \aj, 123, 2183
\bibitem[Songaila(2004)]{song04} Songaila, A.\ 2004, AJ, 127, 
2598
\bibitem[Spergel et al (2003)]{spergeletal} Spergel, D.N., et al,
  2003, ApJS, 148, 175
\bibitem[Stern et al (2004)]{stern} Stern, D., et al, ApJ, submitted, astro-ph/0407409
\bibitem[Telfer et al.(2002)]{telfer} Telfer, R. C., Zheng, W., Kriss, G. A. 2002, \aj, 565, 773
\bibitem[White et al.(2003)]{white03} White, R. L., Becker, R. H., Fan, X., \& Strauss, M. A. 2003, \aj, 126, 1
\bibitem[Wyithe \& Loeb(2004a)]{wl04a} Wyithe, J. S. B., \& Loeb, A. 2004a, \nat, 427, 815
\bibitem[Wyithe \& Loeb(2004b)]{wl04b} Wyithe, J. S. B., \& Loeb, A. 2004b, \nat, in press (astro-ph/0409412)

\end{thebibliography}
\end{document}